\documentclass[utf8]{frontiersSCNS} % for Science, Engineering and Humanities and Social Sciences articles
\usepackage{url,hyperref,lineno,microtype,subcaption}
\usepackage[onehalfspacing]{setspace}

\def\keyFont{\fontsize{8}{11}\helveticabold }
\def\firstAuthorLast{Trakhtenbrot {et~al.}} %use et al only if is more than 1 author
\def\Authors{Benny Trakhtenbrot\,$^{1,*}$, Paulina Lira\,$^{2}$, Hagai Netzer\,$^{3}$, Claudia Cicone\,$^{1,4}$, Roberto Maiolino\,$^{5,6}$ and Ohad Shemmer\,$^{7}$}
\def\Address{$^{1}$Department of Physics, ETH Zurich, Wolfgang-Pauli-Strasse 27, CH-8093 Zurich, Switzerland \\
$^{2}$Departamento de Astronomia, Universidad de Chile, Camino del Observatorio 1515, Santiago, Chile \\  
$^{3}$School of Physics and Astronomy and the Wise Observatory, The Raymond and Beverly Sackler Faculty of Exact Sciences, Tel-Aviv University, Tel-Aviv 69978, Israel \\  
$^{4}$INAF-Osservatorio Astronomico di Brera, via Brera 28, I-20121, Milano, Italy \\  
$^{5}$Cavendish Laboratory, University of Cambridge, 19 J. J. Thomson Avenue, Cambridge CB3 0HE, UK \\  
$^{6}$Kavli Institute for Cosmology, University of Cambridge, Madingley Road, Cambridge CB3 0HA, UK \\  
$^{7}$Department of Physics, University of North Texas, Denton, TX 76203, USA 
}
\def\corrAuthor{Benny Trakhtenbrot}

\def\corrEmail{benny.trakhtenbrot@phys.ethz.ch}

\newcommand{\kpc}	{\ifmmode {\rm kpc} \else kpc\fi}
\newcommand{\kms}	{\ifmmode {\rm km\,s}^{-1} \else km\,s$^{-1}$\fi}
\newcommand{\ergs}	{\ifmmode {\rm erg\,s}^{-1} \else erg s$^{-1}$\fi}
\newcommand{\mic}	{\ifmmode {\rm \mu m} \else $\mu$m\fi}
\newcommand{\Hubble}	{\ifmmode {\rm km\,s}^{-1}\,{\rm Mpc}^{-1} \else km\,s$^{-1}$\,Mpc$^{-1}$\fi}
\newcommand{\Msun}{\ifmmode M_{\odot} \else $M_{\odot}$\fi}
\newcommand{\Lsun}{\ifmmode L_{\odot} \else $L_{\odot}$\fi}
\newcommand{\Zsun}{\ifmmode Z_{\odot} \else $Z_{\odot}$\fi}
\newcommand{\Rsun}{\ifmmode R_{\odot} \else $R_{\odot}$\fi}
\newcommand{\mpyr}{\ifmmode \Msun\,{\rm yr}^{-1} \else $\Msun\,{\rm yr}^{-1}$\fi}
\newcommand{\Msol}{\Msun}

\newcommand{  \CII	}{\ifmmode \left[{\rm C}\,\textsc{ii}\right]\,\lambda157.74\,\mu{\rm m} \else [C\,{\sc ii}]\ $\lambda157.74\,\mu{\rm m}$\fi}
\newcommand{  \cii	}{\ifmmode \left[{\rm C}\,\textsc{ii}\right] \else [C\,{\sc ii}]\fi}
\newcommand{  \mgii     }{\ifmmode {\rm Mg}\,\textsc{ii} \else Mg\,\textsc{ii}\fi}
\newcommand{  \MgII     }{\ifmmode {\rm Mg}\,\textsc{ii}\,\lambda2798 \else Mg\,\textsc{ii}\,$\lambda2798$\fi}
\newcommand{ \fwhm  }{\ifmmode {\rm FWHM} \else FWHM\fi} 
\newcommand{ \voff  }{\ifmmode v_{\rm off} \else $v_{\rm off}$\fi} 
\newcommand{ \vmax  }{\ifmmode v_{\rm max} \else $v_{\rm max}$\fi} 
\newcommand{  \mbh      }{\ifmmode M_{\rm BH} \else $M_{\rm BH}$\fi}
\newcommand{  \lledd    }{\ifmmode L/L_{\rm Edd} \else $L/L_{\rm Edd}$\fi}
\newcommand{  \Lbol     }{\ifmmode L_{\rm bol} \else $L_{\rm bol}$\fi}
\newcommand{  \mstar    }{\ifmmode M_{*} \else $M_{*}$\fi} 
\newcommand{  \mgal     }{\ifmmode M_{*} \else $M_{*}$\fi} 
\newcommand{  \mhost    }{\ifmmode M_{\rm host} \else $M_{\rm host}$\fi}
\newcommand{  \mmsmall  }{\ifmmode M_{\rm BH}/M_{*} \else $M_{\rm BH}/M_{*}$\fi}
\newcommand{  \mmlarge  }{\ifmmode M_{*}/M_{\rm BH} \else $M_{*}/M_{\rm BH}$\fi}

\newcommand{  \mdyn     }{\ifmmode M_{\rm dyn} \else $M_{\rm dyn}$\fi} 
\newcommand{  \sfr      }{\ifmmode {\rm SFR} \else SFR\fi}
\newcommand{ \Lcii     }{\ifmmode L_{\cii} \else $L_{\cii}$\fi}
\newcommand{ \fwcii  }{\ifmmode {\rm FWHM}\cii \else FWHM\cii\fi}
\newcommand{  \herschel} {{\it Herschel}}
\newcommand{  \spitzer }  {{\it Spitzer}}
\newcommand{  \hst     }  {{\it HST}}
\newcommand{\zfpe}{$z \simeq 4.8$}

\def\arcsec{\hbox{$^{\prime\prime}$}}

\newcommand\farcs{\mbox{$.\!\!^{\prime\prime}$}}%

\begin{document}
\onecolumn
\firstpage{1}

\title[Quasar Hosts and Mergers at High $z$]{Fast-growing SMBHs in Fast-growing Galaxies, at High Redshifts: the Role of Major Mergers as Revealed by ALMA} 

\author[\firstAuthorLast ]{\Authors} %This field will be automatically populated
\address{} %This field will be automatically populated
\correspondance{} %This field will be automatically populated

\extraAuth{}% If there are more than 1 corresponding author, comment this line and uncomment the next one.
%\extraAuth{corresponding Author2 \\ Laboratory X2, Institute X2, Department X2, Organization X2, Street X2, City X2 , State XX2 (only USA, Canada and Australia), Zip Code2, X2 Country X2, email2@uni2.edu}

\maketitle

\begin{abstract}

%%% Leave the Abstract empty if your article does not require one, please see the Summary Table for full details.
% \section{}
% For full guidelines regarding your manuscript please refer to \href{http://www.frontiersin.org/about/AuthorGuidelines}{Author Guidelines}.

% As a primary goal, the abstract should render the general significance and conceptual advance of the work clearly accessible to a broad readership. References should not be cited in the abstract. Leave the Abstract empty if your article does not require one, please see \href{http://www.frontiersin.org/about/AuthorGuidelines#SummaryTable}{Summary Table} for details according to article type. 

We present a long-term, multi-wavelength project to understand the epoch of fastest growth of the most massive black holes by using a sample of 40 luminous quasars at $z\simeq4.8$.
These quasars have rather uniform properties, with typical accretion rates and black hole masses
of $\lledd\simeq0.7$ and $\mbh\simeq10^9\,\Msun$. 
The sample consists of ``FIR-bright'' sources with a previous \herschel/SPIRE detection, suggesting ${\rm SFR} >1000\,\mpyr$, as well as of ``FIR-faint'' sources for which \herschel\ stacking analysis implies a typical SFR of $\sim400\,\mpyr$. 
Six of the quasars have been observed by ALMA in \CII\ line emission and adjacent rest-frame $150\,\mu {\rm m}$ continuum, to study the dusty cold ISM. 
ALMA detected companion, spectroscopically confirmed sub-mm galaxies (SMGs) for three sources – one FIR-bright and two FIR-faint.
The companions are separated by $\sim 14-45$ kpc from the quasar hosts, and we interpret them as major galaxy interactions. 
Our ALMA data therefore clearly support the idea that major mergers may be important drivers for rapid, early SMBH growth. 
However, the fact that not all high-SFR quasar hosts are accompanied by interacting SMGs, and their ordered gas kinematics observed by ALMA, suggest that other processes may be fueling these systems. 
Our analysis thus demonstrates the diversity of host galaxy properties and gas accretion mechanisms associated with early and rapid SMBH growth.

\tiny
 \keyFont{ \section{Keywords:} sub-millimeter galaxies, galaxy mergers, quasars: supermassive black holes, Quasars: host galaxies, high-redshift galaxies} 
 %All article types: you may provide up to 8 keywords; at least 5 are mandatory.
\end{abstract}

\section{Introduction}
\label{sec:intro}

% For Original Research Articles \citep{conference}, Clinical Trial Articles \citep{article}, and Technology Reports \citep{patent}, the introduction should be succinct, with no subheadings \citep{book}. For Case Reports the Introduction should include symptoms at presentation \citep{chapter}, physical exams and lab results \citep{dataset}.

The highest-redshift quasars, observed at $z\sim5-7$ suggest that supermassive black holes (SMBHs) with $\mbh\simeq10^9\,\Msol$ existed about 1 Gyr after the big bang, which challenges our understanding of BH formation and early growth, and how these processes relate to the galaxies that host the earliest  SMBHs.

In order to account for the observed high BH masses of the earliest quasars, many models have promoted the possibility of high-mass BH seed formation, in dense stellar populations in proto-galaxies and/or through the direct collapse of gaseous halos \cite[see, e.g.,][for reviews]{Natarajan2011_seeds_rev,Volonteri2012_Science_rev}.
Regardless of the seed mass, the subsequent BH growth must proceed at high accretion rates and high duty cycles.
The former is indeed directly observed, as the accretion rate of high-$z$ quasars approaches $\lledd\simeq1$ \cite[e.g.,][]{Kurk2007,Willott2010_MBH,DeRosa2011,Trakhtenbrot2011}.
The latter requirement is found to be somewhat more challenging.
One way to efficiently fuel SMBH accretion is through major mergers of gas rich galaxies \cite[][]{Sanders1988,Hopkins2006}.
Such mergers would be more common in dense large-scale environments.
Moreover, several simulations have suggested that over-dense large-scale environments would expedite the growth of the most massive early BHs, as large amounts of inter galactic gas could stream onto the SMBHs host galaxies \cite[][]{Dekel2009_cold_streams,DiMatteo2012,Dubois2012_hiz_inflows,Costa2014_z6_env_sims}.

Regardless of the exact mechanism driving the nearly continuous SMBH fueling, the low angular momentum gas is expected to trigger intense star formation (SF) throughout the host, and any interacting galaxy.
Several observations of high-redshift quasars (including our own; see \S\ref{sec:sample_and_obs}) have indeed identified intense SF, with growth rates exceeding ${\rm SFR}\sim1000\,\mpyr$ \cite[e.g.,][]{Mor2012_z48,Netzer2014_z48_SFR,Netzer2016_herschel_hiz}.
Although these high SFRs are suggestive of merger activity, the low spatial resolution of the far-IR (FIR) data prohibited any detailed investigation of this possibility.
Other dedicated searches for close companions have identified some examples of major mergers \cite[][]{Wagg2012_z47_QSO}, but most searches did not yield convincing evidence for merger activity \cite[e.g.,][]{Willott2005_z6_comp}.
Similarly, wide-field imaging campaigns aimed at determining whether high-$z$ quasars are found in over-dense large-scale environments yielded ambiguous results \cite[][]{Willott2005_z6_comp,Kim2009_idrops_z6,Husband2013,Banados2013_z57_env,Simpson2014_ULASJ1120_env}.

Here we describe a pilot study with ALMA that aims to identify major galaxy-galaxy interactions among a sample of six fast-growing SMBHs at \zfpe.
The full presentation of this study was recently published in \emph{The Astrophysical Journal} \cite[][T17 hereafter]{Trakhtenbrot2017_z48_ALMA}, and here we only provide a brief summary of the sample, the ALMA observations, and our main results. 
The interested reader is encouraged to refer to T17 for any additional details.
Throughout this work, we assume a cosmological model with $\Omega_{\Lambda}=0.7$, $\Omega_{\rm M}=0.3$, and $H_{0}=70\,\kms\,{\rm Mpc}^{-1}$.

\section{Sample and ALMA Observations}
\label{sec:sample_and_obs}

% For requirements for a specific article type please refer to the Article Types on any Frontiers journal page. Please also refer to  \href{http://home.frontiersin.org/about/author-guidelines#Sections}{Author Guidelines} for further information on how to organize your manuscript in the required sections or their equivalents for your field

% For Original Research articles, please note that the Material and Methods section can be placed in any of the following ways: before Results, before Discussion or after Discussion.

Our sample of six quasars is drawn from a larger sample of 40 sources at \zfpe, for which reliable estimates of \mbh, \lledd, and integrated host SFRs are available through our long-term, multi-wavelength observational effort, conducted using the VLT, Gemini, \spitzer, and \herschel\ facilities.
The \zfpe\ quasars typically have $\mbh\simeq10^{9}\,\Msun$ and $\lledd\simeq0.7$, and the sample covers a rather limited range in these two quantities \cite[see][T11 hereafter]{Trakhtenbrot2011}.
The host galaxies, on the other hand, exhibit a wide range in SFRs. 
While $\sim$75\% of the systems have $\sfr\sim400\,\mpyr$, as determined from \herschel\ stacking analysis (``FIR-faint'' systems), the outstanding 25\% are individually detected and have $\sfr\sim1000-4000\,\mpyr$ \cite[``FIR-bright'' systems; see][]{Netzer2014_z48_SFR,Netzer2016_herschel_hiz}.
The \herschel\ data available prior to the ALMA campaign is therefore suggestive of a scenario where major mergers may be in play in at least in a fraction of these systems.
The six quasars selected for our pilot ALMA study are equally split between ``FIR-bright'' and ``FIR-faint'' subsets, in an attempt to address this possibility.

The ALMA band-7 observations were designed to detect and resolve, at kpc scales, the emission from the prominent \CII\ line and the adjacent continuum.
While the continuum emission probes the spatial distribution of cold dusty ISM in the quasar hosts, the \cii\ emission line -- which is an efficient ISM coolant -- probes their kinematics and can be used to spectroscopically confirm the nature of any companion galaxies \cite[e.g.,][]{Maiolino2009_CII_hiz,Wagg2012_z47_QSO,Wang2013_z6_ALMA,Neri2014_CII_HDF850.1}. 
We used the extended C34-4 configuration of ALMA, providing a resolution of $\sim$0\farcs3 at 330 GHz. 
This corresponds to about 2 kpc at \zfpe. 
The ALMA field of view covers distances of $\sim$6\farcs8, or almost 50 \kpc, from the quasar locations.
The chosen spectral setup provided four windows, each covering 1875 MHz ($\sim$1650 \kms), at a resolution of $\sim$30 \kms.
On-source integrations lasted between 11-54 minutes, with longer integrations for the ``FIR-faint'' sources. 
The resulting limiting flux densities were $F_\nu \sim (4.2-9.2)\times10^{-2}$ mJy/beam (rms).
At the redshifts of the quasars, and under reasonable assumptions regarding the possible shapes of their FIR SEDs, this corresponds to lower limits of roughly $4-11\,\mpyr\,\kpc^{-2}$ (at the $3\sigma$ level).

% We followed standard reduction procedures, using the CASA package (version 4.5.0; \citet{Mcmullin2007_CASA}), and applying the \texttt{CLEAN} algorithm with different weighting setups.

\section{Results}
\label{sec:results}

\begin{figure}[h!]
\begin{center}
\includegraphics[width=0.475\textwidth]{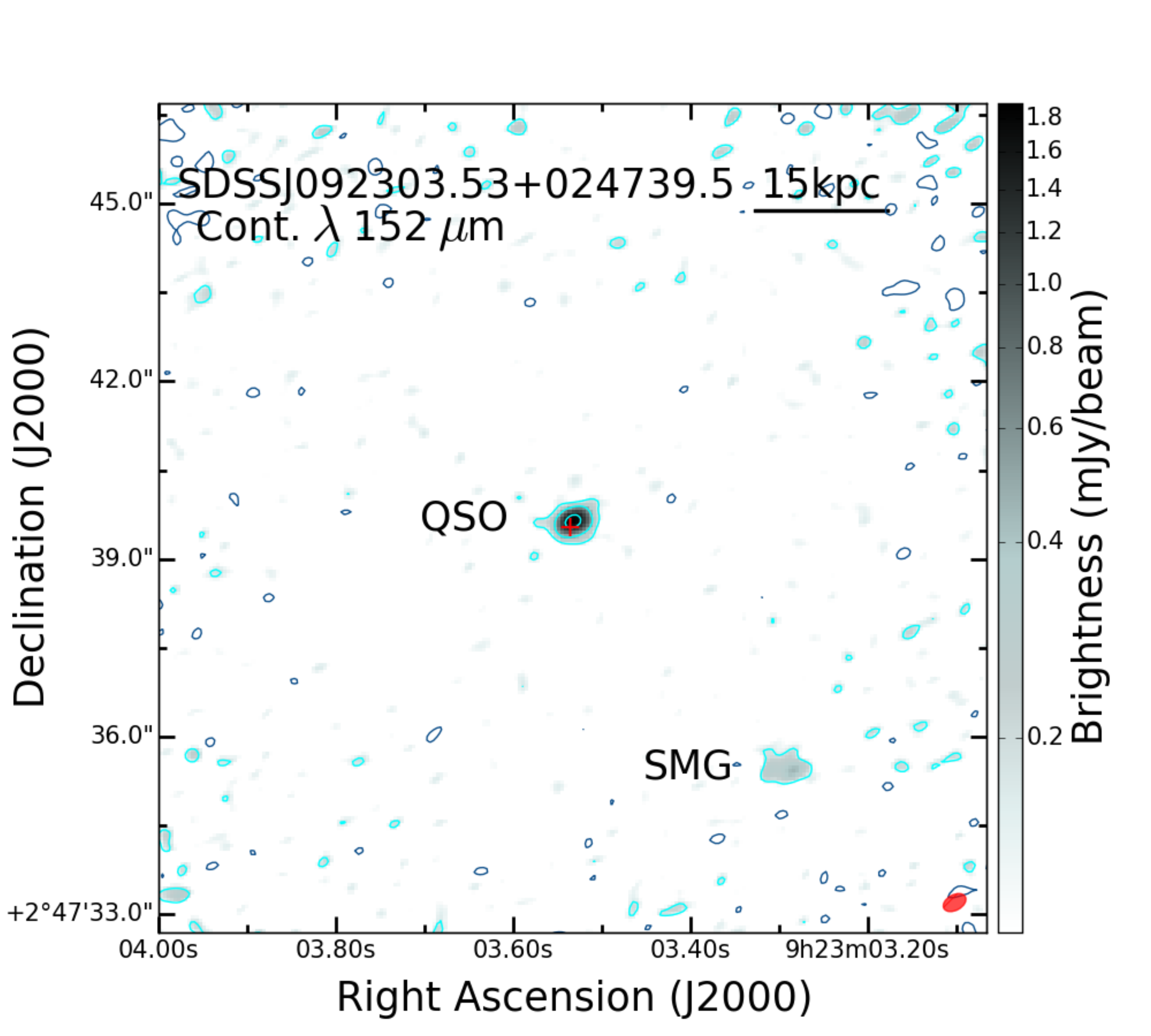}% This is a *.jpg file
\hfill
\includegraphics[width=0.475\textwidth]{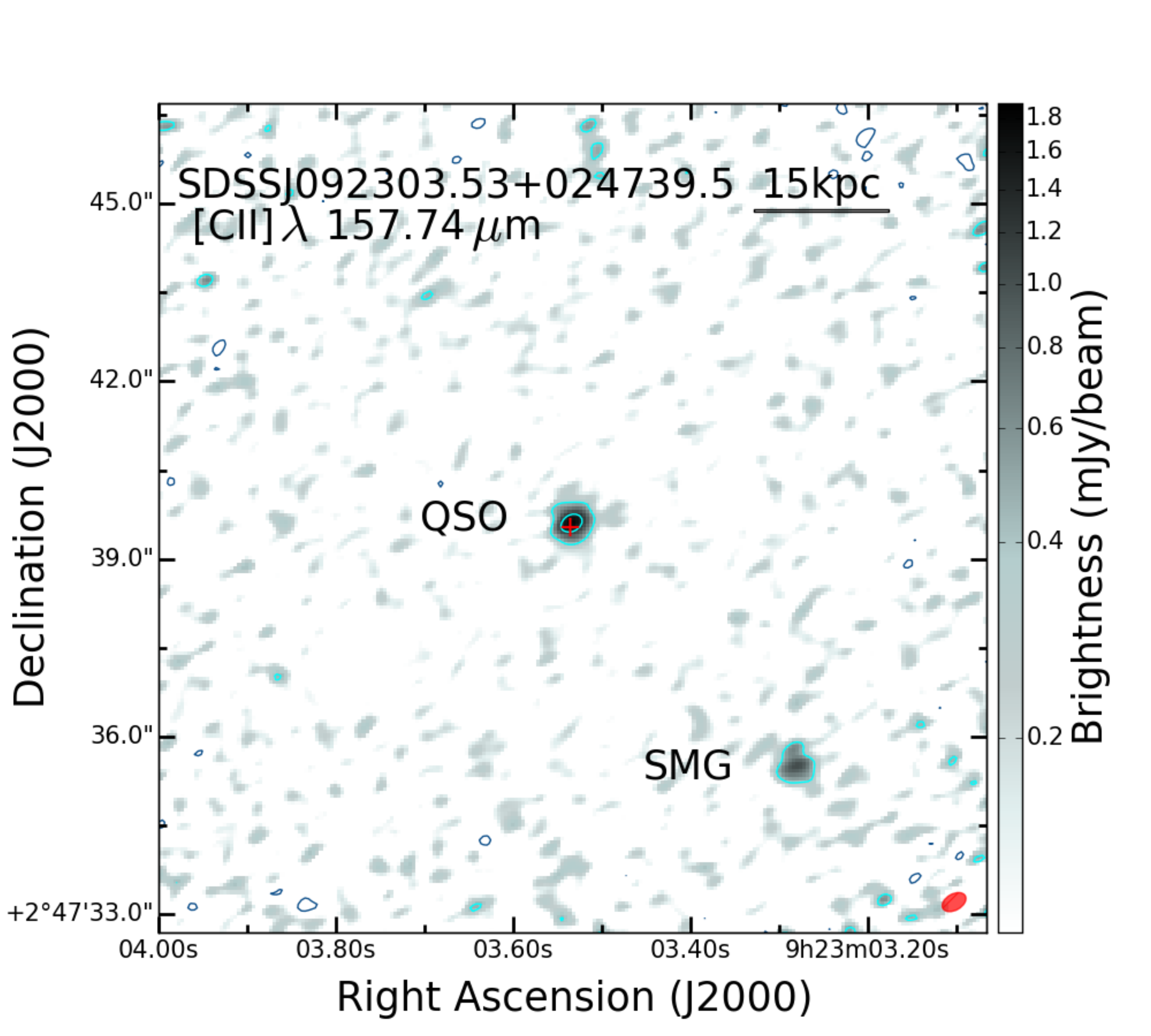}% This is a *.jpg file
\end{center}
\caption{
Maps of the dust continuum (\emph{left}) and \CII\ line (\emph{right}) emission for one of the quasars in our sample, SDSS~J092303.53+024739.5 at $z_{\rm QSO}=4.6589$.
Both the quasar host (marked as ``QSO'') and a sub-mm companion galaxy (``SMG'') are robustly detected and marginally resolved (see the synthetized ALMA beam at the bottom right of each panel).
The companion galaxy is separated by 36.5 kpc and 246 \kms\ from the quasar host.
}
\label{fig:maps_cont_cii}
\end{figure}

The host galaxies of all six quasars are robustly detected, and (marginally) resolved, in both continuum and \cii\ emission.
As an example, we show inf Fig.~\ref{fig:maps_cont_cii} the continuum and \cii\ emission maps of one of the ``FIR-faint'' sources in our sample, SDSS J092303.53+024739.5 ($z_{\rm QSO}=4.6589$; J0923 hereafter).

In what follows, we highlight our main findings from the analysis of these data. 
We demonstrate these findings using different diagrams for the aforementioned source J0923.
We note that many of the choices we made through the analysis of the ALMA data were motivated by recent sub-mm studies of $z\gtrsim5$ quasars \cite[][]{Wang2013_z6_ALMA,Willott2015_CFHQS_ALMA,Venemans2016_z6_cii}.
The reader is referred to T17 for a detailed discussion of our analysis and assumptions.

\subsection{Quasar hosts}
\label{subsec:hosts}

% \subsubsection{continuum, SFR, agreement with \herschel}

\begin{figure}[h!]
\begin{center}
\includegraphics[clip,trim={0 0 0.7cm 0},width=0.475\textwidth]{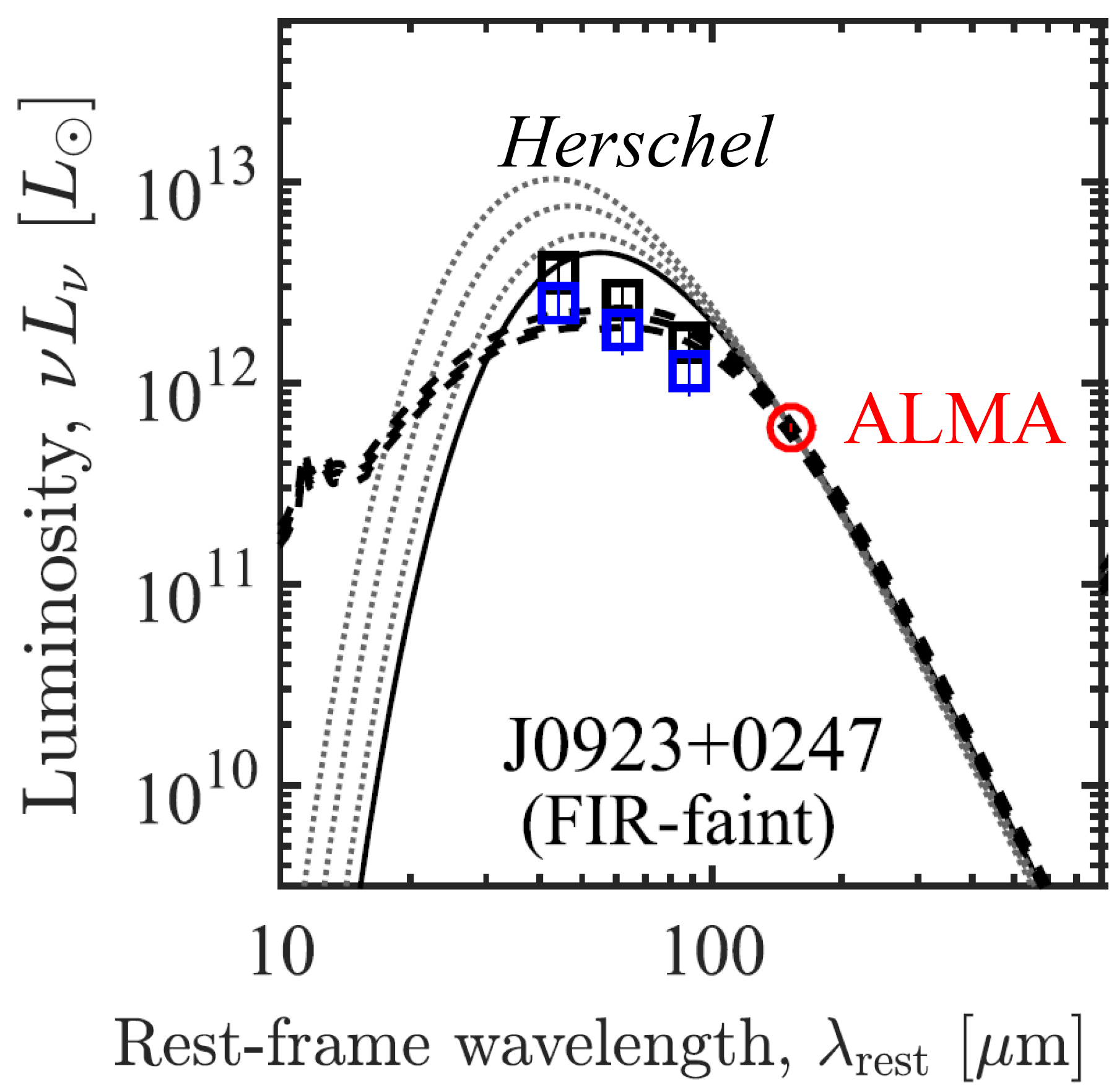}% This is a *.jpg file
\end{center}
\caption{The FIR SED of J0923.
The new ALMA continuum measurement (red) is broadly consistent with the previous \herschel\ data (based on stacking analysis; black squares). 
Blue squares show the \herschel\ data after correcting for the fraction of the flux that comes from the companion galaxy. 
The solid black line traces a gray-body SED with $T_{\rm d}=47$ K and $\beta=1.6$, while the dotted lines trace different temperatures.
The dashed lines illustrate several relevant templates from \cite{CharyElbaz2001_FIR}.
}
\label{fig:fir_sed}
\end{figure}

% The continuum emission from the quasar hosts has spatially integrated flux densities covering a wide range, $F_\nu \ simeq 1.6-18.5$ mJy (at roughly 345 GHz). 
We measure a wide range in (spatially-integrated) 345 GHz continuum flux densities, between $F_\nu\simeq 1.6-18.5$ mJy. 
This wide range in continuum levels is reminiscent of that of the FIR luminosities and SFRs measured from the \herschel/SPIRE data (which covered rest-frame wavelengths of $\sim$45-90 \mic).
Indeed, we find that the new ALMA continuum measurements are generally in very good agreement with the \herschel\ measurements, under reasonable assumptions regarding the shape of the FIR SED, namely a gray-body with dust temperature $T_{\rm d}=47$ K and $\beta = 1.6$. 
Some sources require somewhat warmer dust temperatures (up to $T_{\rm d}\simeq60$ K).
Moreover, most sources are consistent with the FIR SED templates of \cite{CharyElbaz2001_FIR}.
Importantly, we note that the ALMA continuum measurements for the FIR-faint sources, are consistent with the extrapolation of the \emph{stacking} measurements of the \herschel\ data, thus reassuring that our interpretation of the \herschel\ results was robust.
Figure~\ref{fig:fir_sed} demonstrates these findings for J0923.

By combining the new ALMA continuum measurements and the assumed FIR SEDs, we estimate (spatially integrated) total FIR luminosities of $L(8-1000\,\mic)\simeq(1.9-35.5)\times10^{12}\,\Lsun$.
These luminosities translate to host SFRs in the range $\sfr\sim 190-3500\,\mpyr$. 
This is, again, consistent with our \herschel-based findings, but now robustly resolving the hosts, which is crucial in several cases (see \S\ref{subsec:companions} below).

% \subsubsection{dynamical masses}

\begin{figure}[h!]
\begin{center}
\includegraphics[width=0.475\textwidth]{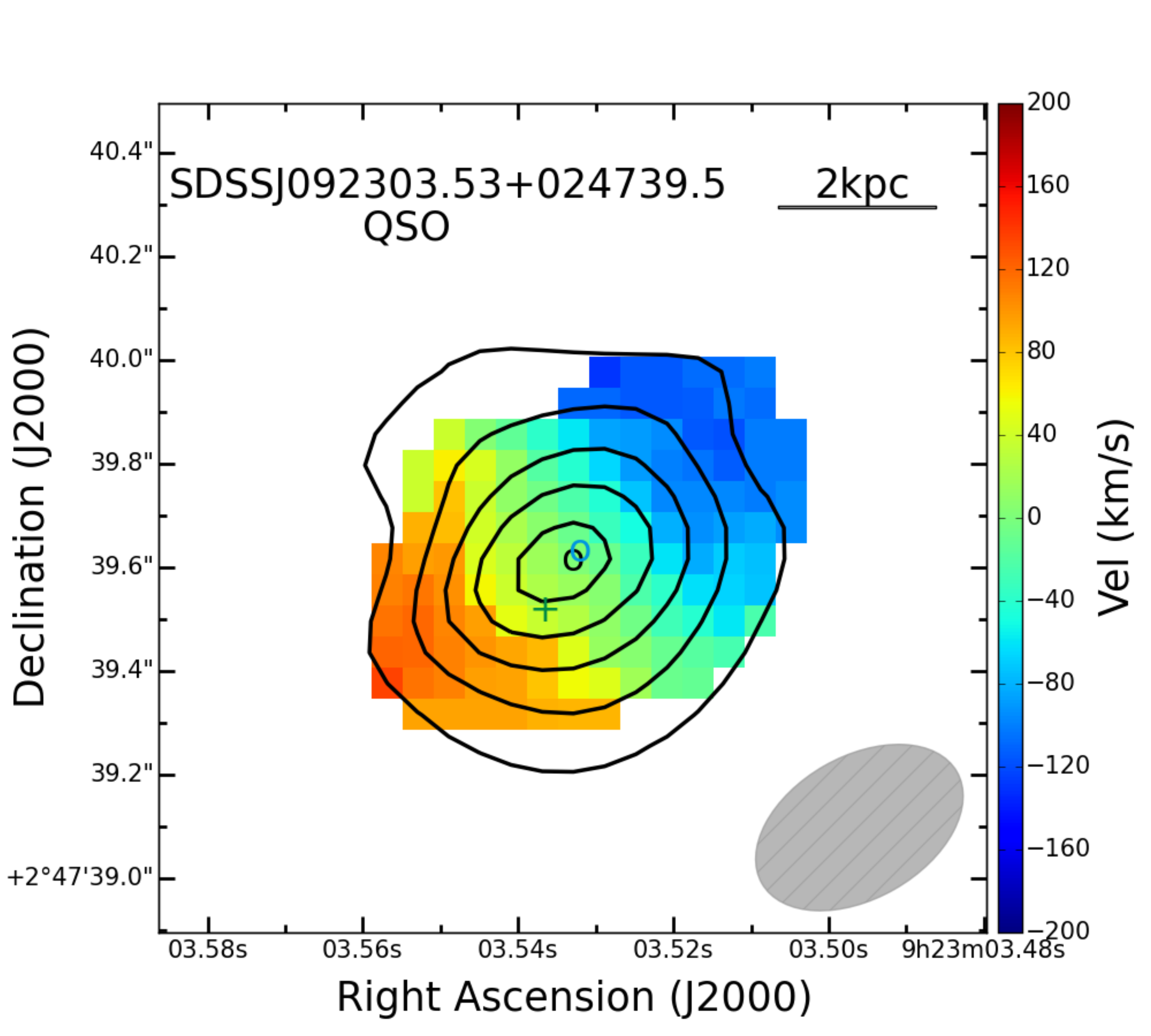}% This is a *.jpg file
\hfill
\includegraphics[width=0.475\textwidth]{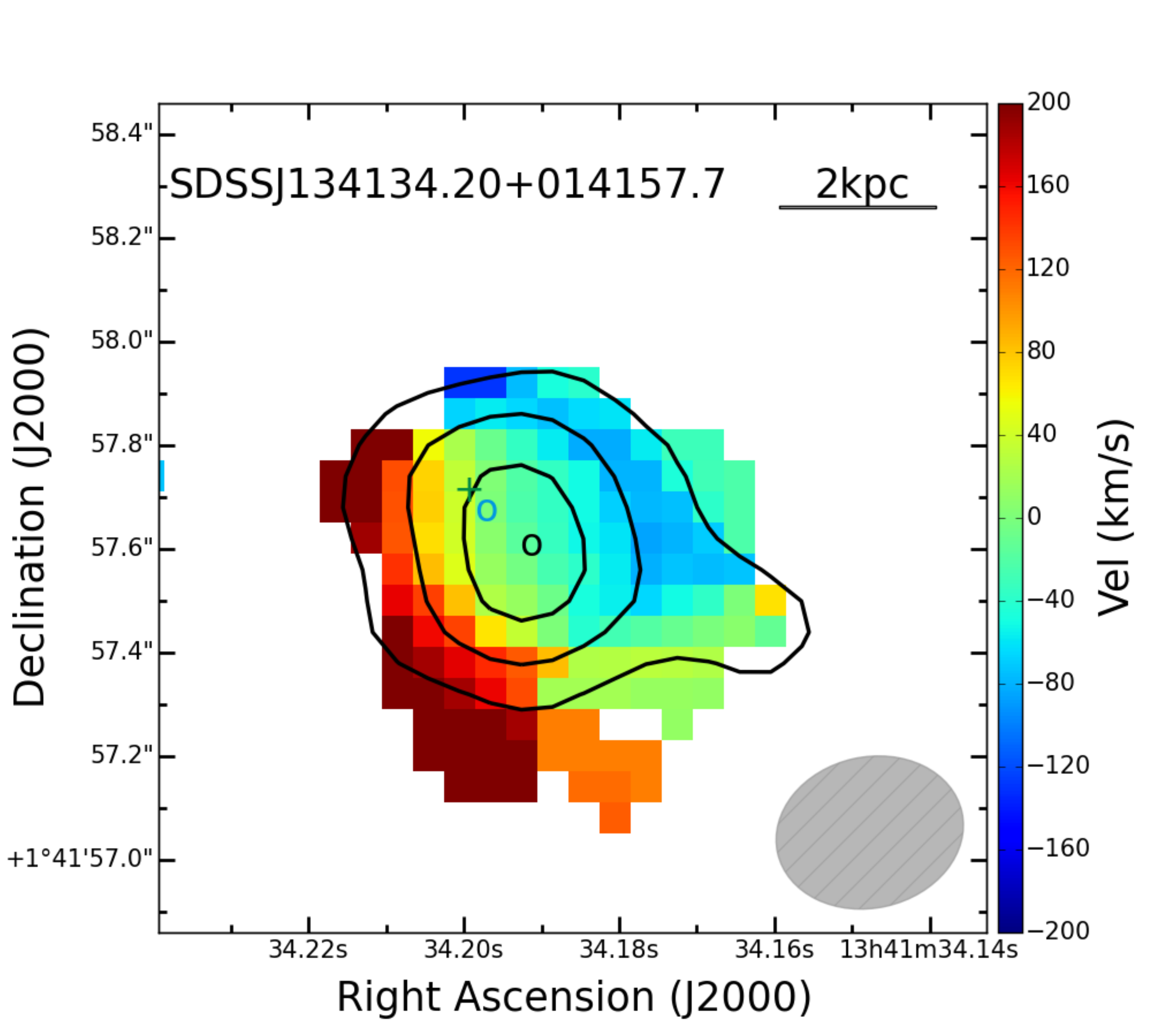}% This is a *.jpg file
\end{center}
\caption{Velocity maps of the \CII\ emission line for J0923 (left) and J1341 (right), with the velocities indicated by color and the \cii\ flux contours overlaid (note that the contour levels differ between the two sources).
The kinematics of the cold gas in the host galaxies appears to be dominated by rotation.}
\label{fig:cii_map}
\end{figure}

The spatially resolved \cii\ line emission maps allow us to study the kinematics of the hosts, and estimate their dynamical masses.
%
% Most systems show \cii\ velocity maps that are dominated by rotation, as demonstrated by J0923 in the left panel of Fig.~\ref{fig:cii_map}.
Most sources (at least four out of six) show \cii\ velocity gradients that are consistent with rotation, as shown in the left panel of Fig.~\ref{fig:cii_map} for J0923.
We therefore assume a simple model of an inclined rotating disk for the \cii-emitting ISM in the hosts.
Following common practices with similar data, we can then deduce \emph{dynamical} host masses, by combining the size of the \cii-emitting region ($D_{\cii}$) with the typical velocity of the gas ($\fwcii$), corrected for the inclination of the disk ($i$):
\begin{equation}
\mdyn = 9.8 \times 10^{8} \left(\frac{D_{\cii}}{\kpc}\right) \left[\frac{\fwcii}{100\,\,\kms}\right]^2\,\sin^{-2}\left(i\right) \,\,\Msol \,\, .
\label{eq:mdyn_fwcii}
\end{equation}
The inclination of each system is estimated from the spatial shape (morphology) of the \cii\ emitting region, available from our resolved ALMA data (i.e., the major-to-minor axis ratio).

The resulting dynamical masses cover a rather limited range, $\mdyn\simeq (3.7-7.4)\times10^{10}\,\Msol$.
By assuming that the dynamical masses are dominated by the stellar components, and considering the wide range in SFRs, this means that the lower-SFR (FIR-faint) hosts are consistent with the so-called ``main sequence'' of SF galaxies \cite[e.g.,][and references therein]{Speagle2014,Steinhardt2014_SPLASH_MS}, while the high-SFR (FIR-bright) hosts would lie above it. 
Moreover, given the narrow range in \mbh\ and \lledd\ of the quasars themselves, it appears that these host properties are not directly linked to the SMBH properties.

\subsection{Companion galaxies}
\label{subsec:companions}

Our most intriguing finding is related to the detection of several gas rich companions, which are likely interacting with the quasar hosts.
% companion, possibly interacting sub-mm galaxies in the vicinity of the quasar hosts.
% , which were not previously known from our multi-wavelength data.

% \subsubsection{identification, distances, offsets}

\begin{figure}[h!]
\begin{center}
\includegraphics[width=0.475\textwidth]{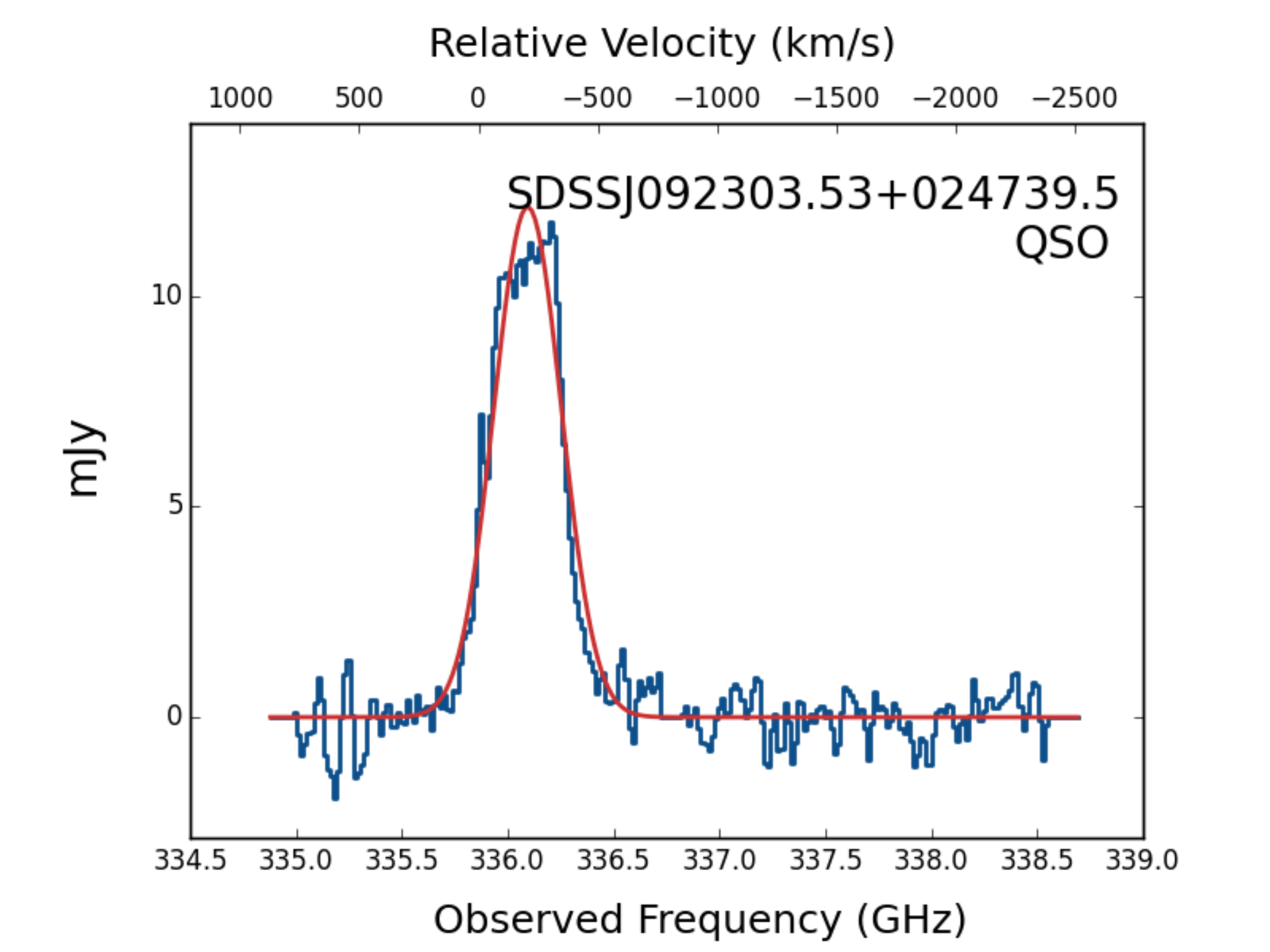}% This is a *.jpg file
\hfill
\includegraphics[width=0.475\textwidth]{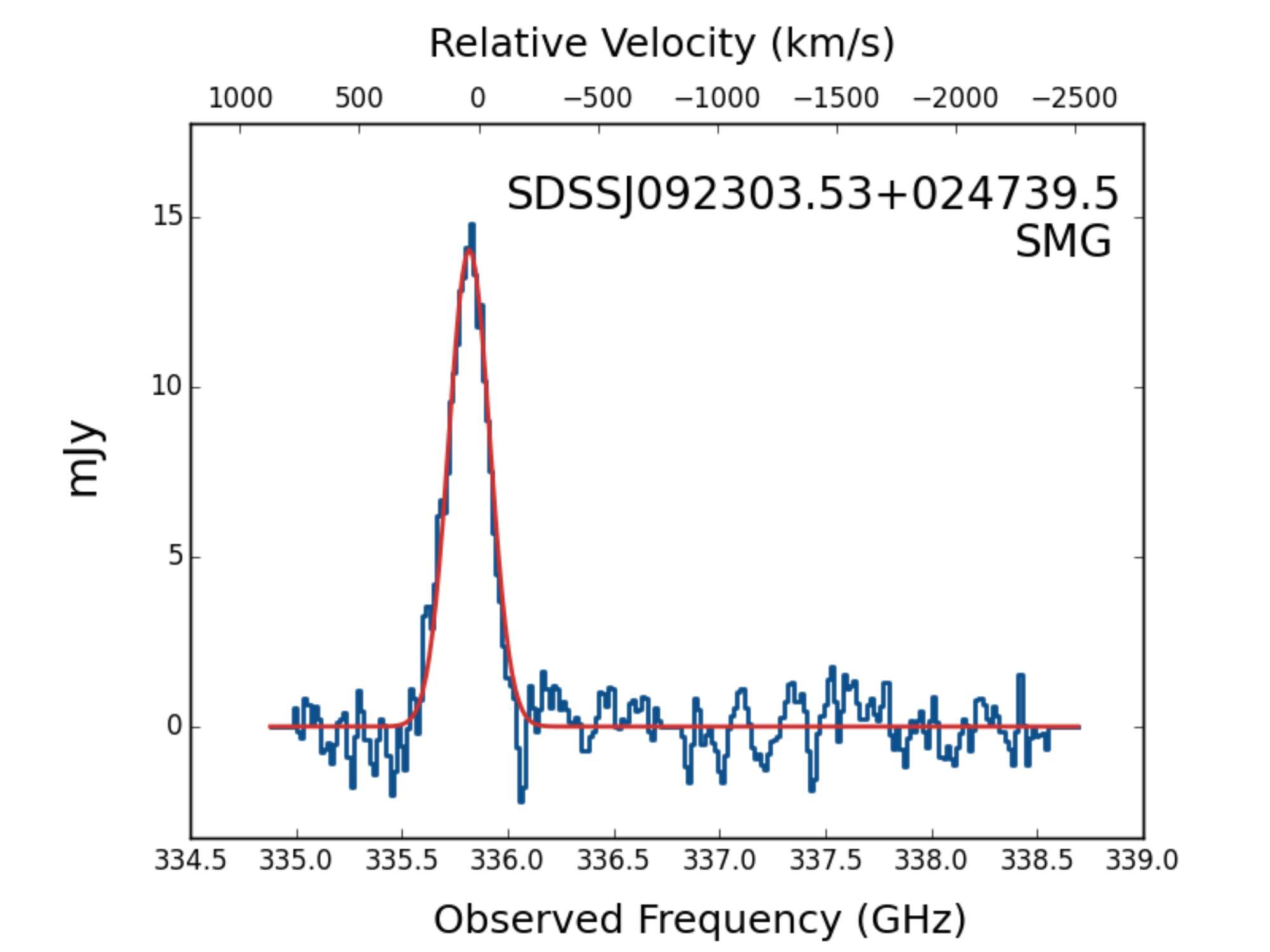}% This is a *.jpg file
\end{center}
\caption{
Spatially-integrated spectra of the \cii\ emission line in the quasar host (\emph{left}) and the interacting companion (\emph{right}) of J0923.
The relative velocities (top $x$-axis) are calculated relative to the rest-frame UV \MgII\ broad emission line of the quasar.
These data confirm the physical association of the companion source to the quasar host.
}
\label{fig:spec_comp}
\end{figure}

We robustly detect companion galaxies for three of the six quasar hosts, in both continuum and \cii\ emission.
These companions are separated by $\sim$14-45 kpc and $|\Delta v|<450\,\kms$ from the quasar hosts, thus being truly physically related to the quasars systems.
An additional continuum source that lacks \cii\ emission is detected $\sim$25 kpc away from one of the FIR-bright systems, which also has a more distant spectroscopically-confirmed companion.
Fig.~\ref{fig:spec_comp} demonstrates the spectral proximity of the companion of J0923 to the quasar host.

Following the same procedures as those used for the quasar hosts, we find that the companion galaxies have continuum fluxes that translate to SFRs of $\simeq100-200\,\mpyr$, and dynamical masses of $\mdyn\simeq(2.1-10.7)\times10^{10}\,\Msol$.
Compared to the respective quasar host masses, the companions have mass ratios $|q| \lesssim 2:1$, suggestive of \emph{major} galaxy interactions. 
Moreover, the companion galaxies are consistent with being on the main sequence of SF galaxies.

% \subsubsection{stress bright/faint ``surprise''}

% \subsubsection{how many, comp. with surveys}

\section{Discussion and Conclusion}

The most intriguing finding of our ALMA study is the identification of spectroscopically-confirmed companion galaxies for three out of the six quasar hosts in our sample.
Considering the small field of view (FoV) of our ALMA data ($\sim$13.5\arcsec or $\sim$100 kpc in diameter), the number of sub-mm bright galaxies we find is much higher than what is found in ``blind'' surveys. 
For example, surveys of rest-frame UV selected SF galaxies predict roughly 0.01 galaxies with ${\rm SFR}\simeq100\,\mpyr$ in a single ALMA FoV \cite[e.g.,][]{Bouwens2015_hiz_LF,Stark2016_hiz_gals_rev}. 
Even more complete surveys of \cii-emitting galaxies at $z\gtrsim5$ predict of about 0.05 galaxies per each of our ALMA pointings \cite[e.g.,][]{Aravena2016_HUDF_cii}.
We therefore conclude that fast-growing $z\sim5$ SMBHs reside in over-dense environments in the early universe, and that their fast accumulation of mass may be related to enhanced major-merger activity.
Further support for this scenario was recently presented in a large ALMA study of $z\sim6$ quasars, using identical methods to those we used in our study \cite[][]{Decarli2017_z6_ALMA}.
%

% We have confirmed that the hosts of fast-growing SMBHs at $z\sim5$ are forming stars at high rates, of order 100s to 1000s \mpyr. 
% The new ALMA data confirms that the extremely high FIR luminosities of the ``FIR-bright'' sources are generally \emph{not} due to blending of multiple sources. 
The naive expectation from the previously available \herschel\ data would be that the high-SFR (FIR-bright) systems would be associated with major mergers, while the lower-SFR (FIR-faint) systems would show no signs of interaction.
Our ALMA data show a very different picture.
Two of the three companions are found near FIR-faint systems, and only one is associated with a FIR-bright system. 
Conversely, two of the three FIR-bright systems in our sample are not associated with companion, interacting galaxies.
Although in principle these quasar hosts may be in an advanced merger stage (which would remain unresolved in our data), the signatures of rotationally-dominated gas structures would not support this scenario..
This is exemplified in the system J1341, which has ${\rm SFR}\simeq3000\,\mpyr$, and shows signatures of rotation-dominated gas and no companion galaxies (see Fig.~\ref{fig:cii_map}, right).
The two lower-SFR systems with companion galaxies are expected to experience a later increase in SFR. 
This means that the low SFRs we deduced for the ``FIR-faint'' T11 \zfpe\ systems cannot be simply due to the onset of ``AGN feedback'' in the final stages of an episode of SMBH and host growth.
However, a larger sample is needed to clarify which of all these processes dominates the growth of the general $z\sim5$ SMBH population.

The companion galaxies detected with ALMA were \emph{not} seen in our \spitzer\ data.
Given their SFRs and (dynamical) masses, and what is known about the population of rest-frame UV selected SF galaxies at $z\simeq4.8$ \cite[e.g.,][]{Steinhardt2014_SPLASH_MS,Stark2016_hiz_gals_rev}, we conclude that this is due to significant dust obscuration.
This may explain the fact that many previous studies were unable to identify companions and/or over-dense environments for $z\gtrsim5$ quasars. 
High resolution, spectroscopic sub-mm observations are therefore crucial for the study of mergers and environments among the highest-redshift quasars.

We are currently leading an ALMA cycle-4 program that would provide similar data for a dozen additional \zfpe\ quasars from the T11 sample, bringing the total number of such quasars with resolved host ISM kinematics, and close companion mapping, to 18. 
Analysis of the ALMA data for the 12 additional sources is ongoing.
Moreover, we were recently awarded \hst/WFC3/IR time to map the \emph{stellar} component in the host galaxies, and in the close companions of the six quasars described here.
The \hst\ data will also probe the larger-scale environments of the quasars, out to $\sim$400 kpc, allowing us to
detect any additional (unobscured) companions that may be present in this field.

\section*{Author Contributions}

% The Author Contributions section is mandatory for all articles, including articles by sole authors. If an appropriate statement is not provided on submission, a standard one will be inserted during the production process. The Author Contributions statement must describe the contributions of individual authors referred to by their initials and, in doing so, all authors agree to be accountable for the content of the work. Please see  \href{http://home.frontiersin.org/about/author-guidelines#AuthorandContributors}{here} for full authorship criteria.

B.T. led the interpretation of the ALMA data, the preparation of the paper describing our results (T17), and presented the results as a contributed talk in the ``Quasars at All Cosmic Epochs'' meeting.
P.L. was the PI of the ALMA proposal and led the data analysis.
All authors participated in different aspects of the analysis, interpretation, and preparation of this study for publication .

\section*{Funding}
% Details of all funding sources should be provided, including grant numbers if applicable. Please ensure to add all necessary funding information, as after publication this is no longer possible.
H.N.\ acknowledges support by the Israel Science Foundation grant 284/13.
C.C.\ gratefully acknowledges support from the Swiss National Science Foundation Professorship grant PP00P2\_138979/1. 
C.C.\ also acknowledges funding from the European Union's Horizon 2020 research and innovation programme under the Marie Sklodowska-Curie grant agreement No 664931.
R.M.\ acknowledges support by the Science and Technology Facilities Council (STFC) and the ERC Advanced Grant 695671 ``QUENCH''.

\section*{Acknowledgments}
% This is a short text to acknowledge the contributions of specific colleagues, institutions, or agencies that aided the efforts of the authors.

We are grateful for the organizers of the ``Quasars at All Cosmic Epochs'' meeting for the opportunity to present our results in a stimulating and friendly environment.
We thank K.\ Schawinski, L.\ Mayer, R.\ Teyssier, P.\ Capelo, M.\ Dotti and D.\ Fiacconi for useful discussions during the preparation of the paper describing our results (T17).
The work described here made use of the ALMA data set ADS/JAO.ALMA\#2013.1.01153.S. 
ALMA is a partnership of ESO (representing its member states), NSF (USA) and NINS (Japan), together with NRC (Canada), NSC and ASIAA (Taiwan), and KASI (Republic of Korea), in cooperation with the Republic of Chile. The Joint ALMA Observatory is operated by ESO, AUI/NRAO and NAOJ.

% \section*{Supplemental Data}
%  \href{http://home.frontiersin.org/about/author-guidelines#SupplementaryMaterial}{Supplementary Material} should be uploaded separately on submission, if there are Supplementary Figures, please include the caption in the same file as the figure. LaTeX Supplementary Material templates can be found in the Frontiers LaTeX folder 

\bibliographystyle{frontiersinSCNS_ENG_HUMS} % for Science, Engineering and Humanities and Social Sciences articles, for Humanities and Social Sciences articles please include page numbers in the in-text citations
%\bibliographystyle{frontiersinHLTH&FPHY} % for Health, Physics and Mathematics articles

% \bibliography{test}
\bibliography{/home/seyfert/trakht/DB/print/LATEX_packs/library}

\newpage

% trim={<left> <lower> <right> <upper>}

% \begin{figure}[h!]
% \begin{center}
% \includegraphics[width=10cm]{logo1}% This is a *.jpg file
% \end{center}
% \caption{ Enter the caption for your figure here.  Repeat as  necessary for each of your figures}\label{fig:1}
% \end{figure}

% \begin{figure}[h!]
% \begin{center}
% \includegraphics[width=15cm]{logos}
% \end{center}
% \caption{This is a figure with sub figures, (A) is one logo, (B) is a different logo.}\label{fig:2}
% \end{figure}

%%% If you are submitting a figure with subfigures please combine these into one image file with part labels integrated.
%%% If you don't add the figures in the LaTeX files, please upload them when submitting the article.
%%% Frontiers will add the figures at the end of the provisional pdf automatically
%%% The use of LaTeX coding to draw Diagrams/Figures/Structures should be avoided. They should be external callouts including graphics.

\end{document}